\documentclass[oldstyle]{mttr}

\addtocounter{page}{25}

\udk{531.38}

\title[Типы критических точек гиростата Ковалевской в двойном поле]{ТИПЫ КРИТИЧЕСКИХ ТОЧЕК\\  ГИРОСТАТА КОВАЛЕВСКОЙ В ДВОЙНОМ ПОЛЕ}

\author{М.П.~Харламов, П.Е.~Рябов, А.Ю.~Савушкин, Г.Е.~Смирнов}

\thanks{\hspace{-5mm}{Работа выполнена при финансовой поддержке РФФИ (гранты № 10-01-00043, 10-01-97001).}}

\mtt{year=2011,number=41,received=10.11.11}

\address{Волгоградский филиал РАНХиГС, Россия\\
Финансовый ун-т, Москва, Россия\\
МГУ им.~М.В.~Ломоносова, Москва, Россия}
\email{mharlamov@vags.ru, glebevgen@yandex.ru}

\newcommand{\sgrad}{\mathop{\rm sgrad}\nolimits}
\newcommand{\tr}{\mathop{\rm trace}\nolimits}

\begin{document}

\maketitle

\begin{abstract}
В соответствии с классификацией, принятой в теории интегрируемых по Лиувиллю гамильтоновых систем, вычислены типы критических точек всех рангов интегрального отображения задачи о движении гиростата Ковалевской в двойном силовом поле.
\end{abstract}


\section{Исходные соотношения} Обобщение  случая С.В.\,Ко\-ва\-лев\-ской, найденное благодаря вкладу авторов работ \cite{BogRus2,Yeh1,ReySem}, до сих пор мало изучено. Система, описывающая движение гиростата с условиями типа Ковалевской в двух силовых полях (например, поле силы тяжести и магнитном), является неприводимой интегрируемой гамильтоновой системой с тремя степенями свободы. Соответствующие уравнения Эйлера--Пуассона запишем в виде
\begin{equation}\notag
\begin{array}{c}
2\dot \omega _1  = \omega _2 (\omega _3-\lambda)  +
\beta _3 ,\; 2\dot  \omega _2  =  - \omega _1
(\omega _3-\lambda)  - \alpha _3 ,\;
\dot \omega _3   = \alpha _2  - \beta _1 , \\
\dot \alpha _1   = \alpha _2 \omega _3  - \alpha_3
\omega_2,\qquad \dot \beta _1   = \beta _2 \omega _3
-
\beta_3 \omega_2, \\
\dot \alpha _2   = \alpha _3 \omega _1  - \alpha_1
\omega_3,\qquad \dot \beta _2   = \beta _3 \omega _1
-
\beta_1 \omega_3, \\
\dot \alpha _3   = \alpha _1 \omega _2  - \alpha_2
\omega_1,\qquad \dot \beta _3   = \beta _1 \omega _2
- \beta_2 \omega_1.
\end{array}
\end{equation}
Здесь $\omega_i, \alpha_j, \beta_k$ -- компоненты в подвижных осях вектора $\boldsymbol \omega$ угловой скорости тела и векторов ${\boldsymbol \alpha},{\boldsymbol \beta}$, характеристических для силовых полей, т.е. векторов, направленных по оси симметрии поля, с модулем, включающим в себя всю скалярную характеристику взаимодействия тела и поля. Таким образом, векторы центра масс и магнитного момента поля становятся единичными. В частности, геометрические интегралы приводятся к виду
\begin{equation}\label{nq1_1}
{\boldsymbol \alpha} \cdot {\boldsymbol \alpha}=a^2, \qquad {\boldsymbol
\beta} \cdot {\boldsymbol \beta}=b^2, \qquad {\boldsymbol \alpha} \cdot
{\boldsymbol \beta}=0.
\end{equation}
Их совместный уровень в $\mathbb{R}^9({\boldsymbol \omega},{\boldsymbol \alpha},{\boldsymbol \beta})$ обозначим через $\mathcal{P}^6.$ Меняя при необходимости порядок векторов в подвижной системе отсчета и считая задачу несимметричной, получим $a > b > 0$.
Первые интегралы в инволюции таковы:
\begin{equation}\notag
\begin{array}{l}
\displaystyle{H = \omega _1^2  + \omega _2^2  + \frac{1} {2}
\omega_3^2 -
\alpha _1  - \beta _2,} \\
K = (\omega _1^2  - \omega _2^2  + \alpha _1  - \beta _2 )^2 +
(2\omega _1 \omega _2  + \alpha _2  + \beta _1 )^2 + \\[2mm]
\phantom{K = (}+ 2\lambda[(\omega_3-\lambda) ( \omega_1^2+ \omega_2
^ 2)
+ 2 \omega_1 \alpha_3 + 2 \omega_2 \beta_3 ] , \\
\displaystyle{G =\frac{1} {4} (M_{\alpha}^2+ M_{\beta}^2) + \frac{1}
{2} (\omega_3 - \lambda) M_{\gamma} - b^2 \alpha _1  - a^2 \beta
_2.}
\end{array}
\end{equation}
Здесь обозначено
\begin{gather*}
M_\alpha = 2\omega_1 \alpha_1 + 2\omega_2 \alpha_2 +
(\omega_3+\lambda) \alpha_3 , \qquad M_\beta = 2\omega_1 \beta_1 + 2\omega_2 \beta_2 + (\omega_3+\lambda) \beta_3 , \\
M_\gamma = 2\omega_1 (\alpha_2 \beta_3 - \alpha_3 \beta_2 ) +
2\omega_2 (\alpha_3 \beta_1 - \alpha_1 \beta_3 ) + (\omega_3+\lambda)
(\alpha_1 \beta_2 - \alpha_2 \beta_1 ).
\end{gather*}

Введем интегральное отображение
\begin{equation}\label{nq1_2}
J=H{\times}K{\times}G: \mathcal{P}^6 \rightarrow {\mathbb{R}}^3.
\end{equation}
Множество его критических точек состоит из движений тела специального вида. Во-первых, это положения равновесия (критические точки ранга~0). Их в рассматриваемой системе ровно четыре \cite{ZotKh}. Во-вторых, это так называемые особые периодические движения -- такие замкнутые траектории, которые сами являются орбитами всех трех гамильтоновых полей, определенных общими интегралами $H, K, G$, т.е. орбитами пуассонова действия на шестимерном фазовом пространстве. Они составлены из критических точек ранга 1. В это семейство входят и сепаратрисы, возникающие при бифуркации особых периодических движений при прохождении через неустойчивое положение равновесия. Все решения такого вида найдены и сведены к эллиптическим квадратурам в работе \cite{Kh37}. Критические точки ранга 1 организованы в двумерные многообразия.

Наконец, в критическое множество входят двумерные торы Лиувилля, которые также являются орбитами пуассонова действия, т.е. не получены в результате случайной соизмеримости частот на движении общего вида. Двумерные торы состоят из критических точек ранга 2 и формируют множества, которые почти всюду являются четырехмерными симплектическими многообразиями. Слова ``почти всюду'' здесь существенны, так как топология таких почти многообразий достаточно сложна и до конца не выяснена даже в случае нулевого гиростатического момента, когда соответствующие системы сведены к квадратурам \cite{KhSav,Kh38}. Однако в рассматриваемой задаче имеется их аналитическое описание \cite{KhND07,KhHMJ}. Приведем здесь необходимые понятия и формулы.

Бифуркационная диаграмма отображения \eqref{nq1_2} содержится в объединении следующих (пересекающихся) подмножеств
пространства ${\mathbb{R}}^3(h,k,g)$ \cite{KhND07}:

1) прямых
\begin{equation} \notag
\Pi_-: \left\{ \begin{array}{l} k = (a + b)^2 \\
\displaystyle{g= - \,a b (h-\frac{\lambda^2}{2})}
\end{array} \right., \qquad \Pi_+: \left\{ \begin{array}{l} k = (a - b)^2 \\
\displaystyle{g=  \,a b (h-\frac{\lambda^2}{2})}
\end{array} \right.;
\end{equation}

2) поверхности
\begin{equation} \label{nq1_3}
\Pi_1: \left\{ \begin{array}{l} \displaystyle{k =
p^2+(h-\frac{\lambda^2}{2})^2-4(h-\frac{\lambda^2}{2})s+3s^2-
\frac{p^4-r^4}{4s^2},} \\[3mm]
\displaystyle{g=(h-\frac{\lambda^2}{2}-s)s^2+\frac{p^4-r^4}{4s},
\qquad s \in {\mathbb{R}}\backslash\{0\};}
\end{array} \right.
\end{equation}

3) поверхности
\begin{equation} \label{nq1_4}
\Pi_2: \left\{ \begin{array}{l} \displaystyle{k = -
2\lambda^2(h-\frac{\lambda^2}{2}-2s)-\lambda^4+ \frac{r^4}
{4s^2},} \\[3mm]
\displaystyle{g=\frac{1}{2}p^2(h+\frac{\lambda^2}{2})-\lambda^2 s^2-\frac{r^4}{4s},
\qquad s \in {\mathbb{R}}\backslash\{0\}.}
\end{array} \right.
\end{equation}
Здесь и далее используются обозначения $p^2=a^2+b^2$, $r^2=a^2-b^2$.

Множества $\Pi_{\pm}$ порождены семействами маятниковых движений, в которых ось колебаний ортогональна обоим силовым полям (существование таких движений для динамически симметричного гиростата с произвольным отношением осевого и экваториального моментов инерции в двойном поле впервые указано в \cite{Yeh2}):
\begin{equation}\label{nq1_5}
\begin{array}{c} {\boldsymbol \alpha}\times{\boldsymbol \beta} \equiv  \pm a b{\bf e}_3 , \quad
{\boldsymbol \omega } = {\dot \varphi} {\bf e}_3 , \\
{\boldsymbol \alpha } = a({\bf e}_1 \cos \varphi  - {\bf e}_2 \sin \varphi ),\quad
{\boldsymbol \beta } =  \pm b({\bf e}_1 \sin \varphi + {\bf e}_2 \cos \varphi ), \\
{\ddot \varphi}  =  - (a \pm b)\sin \varphi.
\end{array}
\end{equation}
В соответствии с выбором знака обозначим множество точек такого семейства через $\mathcal{M}_{\pm}$. Это -- два двумерных многообразия, диффеоморфных цилиндру.

Множество критических точек интегрального отображения, лежащих в прообразе поверхности $\Pi_i$, назовем критической подсистемой $\mathcal{M}_i$ $(i=1,2)$. Множества $\mathcal{M}_{1,2}$ почти всюду являются четырехмерными многообразиями. Нетрудно видеть, что семейство $\mathcal{M}_-$ целиком содержится в $\mathcal{M}_1$, а семейство $\mathcal{M}_+$ пересекается с подсистемой $\mathcal{M}_1$ только по движениям со значениями энергии за пределами интервала
\begin{equation}\label{nq1_6}
    \frac{\lambda^2}{2}-2\sqrt{a b} < h < \frac{\lambda^2}{2}+2\sqrt{a b}.
\end{equation}
Пересечения многообразий $\mathcal{M}_{\pm}$ с критической подсистемой $\mathcal{M}_2$ также легко находятся. Это -- движения, лежащие на уровнях $h$, содержащих положения равновесия (по два уровня в каждом семействе). Кроме того, в каждом из семейств имеется по одному уровню $h$, попадающему в $\mathcal{M}_2$, движения на котором оказываются вырожденными как критические точки ранга 1. Соответствующие значения энергии указываются ниже при исследовании типов критических точек.

Введем систему комплексных координат
\begin{equation}\notag
\begin{array}{c}
x_1 = (\alpha_1  - \beta_2) + i(\alpha_2  + \beta_1),\quad
x_2 = (\alpha_1  - \beta_2) - i(\alpha_2  + \beta_1 ), \\
y_1 = (\alpha_1  + \beta_2) + i(\alpha_2  - \beta_1), \quad y_2 =
(\alpha_1  + \beta_2) -
i(\alpha_2  - \beta_1), \\
z_1 = \alpha_3  + i\beta_3, \quad
z_2 = \alpha_3  - i\beta_3,\\
w_1 = \omega_1  + i\omega_2 , \quad w_2 = \omega_1  - i\omega_2,
\quad w_3 = \omega_3.
\end{array}
\end{equation}
Для параметров $p,r$ имеем представление:
\begin{equation}\label{nq1_7}
    p^2=\frac{1}{2}(x_1 x_2+y_1 y_2)+z_1 z_2, \qquad r^4 =(x_1 y_2+z_1^2)(x_2 y_1+z_2^2).
\end{equation}

Описание множеств $\mathcal{M}_1,\mathcal{M}_2$ можно получить следующим образом \cite{KhND07}.
Множество $\mathcal{M}_1$ есть замыкание подмножества в фазовом пространстве $\mathcal{P}^6$, заданного уравнениями
\begin{equation}\label{nq1_8}
\begin{array}{l}
y_1= - \displaystyle{\frac{1}{w_1 w_2 (w_3-\lambda)}}\{w_1(w_3-\lambda)[x_2 w_1+z_2(w_3+\lambda)] +\\[2mm]
\phantom{y_1=}+ x_2 z_1 w_1+x_1 z_2 w_2+z_1 z_2 (w_3+\lambda)\},
\\[2mm]
y_2= - \displaystyle{\frac{1}{w_1 w_2 (w_3-\lambda)}}\{w_2(w_3-\lambda)[x_1 w_2+z_1(w_3+\lambda)] +\\[2mm]
\phantom{y_2=}+ x_2 z_1 w_1+x_1 z_2 w_2+z_1 z_2 (w_3+\lambda)\}.
\end{array}
\end{equation}
При этом фигурирующий в \eqref{nq1_3} параметр $s$ -- постоянная частного интеграла $S_1$ на $\mathcal{M}_1$:
\begin{equation}\label{nq1_9}
\displaystyle{S_1=\frac{x_2 z_1 w_1+x_1 z_2 w_2+z_1 z_2 (w_3+\lambda)}{2\,w_1 w_2 (w_3-\lambda)}.}
\end{equation}

В свою очередь, множество $\mathcal{M}_2$ есть замыкание подмножества в фазовом пространстве $\mathcal{P}^6$, заданного уравнениями
\begin{equation}\label{nq1_10}
\begin{array}{l}
\displaystyle{y_1 = \frac{1}{(w_1 w_2+\lambda w_3)(w_2 x_1+\lambda z_1)\lambda}[w_2(w_1^2+x_1)(x_2 z_1 w_1+x_1 z_2 w_2} - \\[3mm]
\qquad -x_1 x_2 w_3+2 z_1 z_2 \lambda)+x_2(w_1 w_3+z_1)(w_1 z_1-x_1 w_3)\lambda - \\[1.5mm]
\qquad -(x_1 w_3^2-2 z_1 w_1 w_3-z_1^2)z_2 \lambda^2],\\[1.5mm]
\displaystyle{y_2 = \frac{1}{(w_1 w_2+\lambda w_3)(w_1 x_2+\lambda z_2)\lambda} [w_1(w_2^2+x_2)(x_2 z_1 w_1+x_1 z_2 w_2} - \\[3mm]
\qquad -x_1 x_2 w_3+ 2 z_1 z_2 \lambda)+x_1(w_2 w_3+z_2)(w_2 z_2-x_2 w_3)\lambda -\\[1.5mm]
\qquad -(x_2 w_3^2-2 z_2 w_2 w_3-z_2^2)z_1 \lambda^2].
\end{array}
\end{equation}
Параметр $s$ бифуркационной поверхности \eqref{nq1_4} -- постоянная частного интеграла $S_2$ на $\mathcal{M}_2$:
\begin{equation}\notag
\begin{array}{l}
\displaystyle{ S_2= \frac{x_1 x_2 w_3-x_2 z_1 w_1-x_1 z_2 w_2-\lambda z_1 z_2}{2\lambda(w_1 w_2+\lambda w_3)}.}
\end{array}
\end{equation}

Подсистемы $\mathcal{M}_1,\mathcal{M}_2$ пересекаются также по множеству критических точек ранга 1, составляющих семейство особых периодических движений, заданных уравнениями (27), (29), (30) работы \cite{Kh37}. Их для краткости здесь не повторяем. Указанная система уравнений, на которую будем далее ссылаться как на уравнения (ОПД), выражает все фазовые переменные через одну переменную $w$, физический параметр $\lambda$ и два вспомогательных параметра $\sigma, u$, связанных соотношением
\begin{equation} \label{nq1_11}
\begin{array}{l}
\lambda^2(\lambda^2+\sigma)^2 u^5+(\lambda^2+\sigma) [2p^2\lambda^4-(\lambda^2+\sigma)^3 \sigma]\sigma u^4 + \\
\qquad +r^4\lambda^6\sigma^2 u^3+ 2 r^4 \lambda^4 \sigma^4(\lambda^2+\sigma)^2u^2-r^8\lambda^8 \sigma^6=0.
\end{array}
\end{equation}
Изменение единственной свободной переменной $w(t)$ определяется дифференциальным уравнением (32) работы \cite{Kh37}, которое интегрируется в эллиптических функциях.

Отметим, что обращение в нуль знаменателей в выражениях, связанных с подсистемами $\mathcal{M}_1,\mathcal{M}_2$, приводит либо к той части движений вида \eqref{nq1_5}, которая лежит в $\mathcal{M}_1$, либо к движениям ранга 1, описанным системой (ОПД). Условия существования последних подробно исследованы в \cite{KhIISmir}.

Цель настоящей работы -- вычислить в терминах введенных параметров аналитические характеристики, выражающие тип критических точек в смысле определения \cite{BolFom}. Знание типа критической точки интегрируемой системы отвечает и на все вопросы, связанные с характером устойчивости проходящей через нее траектории. Для положений равновесия и маятниковых движений \eqref{nq1_5} тип критических точек был представлен в докладе \cite{RyabConf}. В частности, было показано, что интервал \eqref{nq1_6} для движений семейства $\mathcal{M}_+$ отвечает особенности типа ``фокус-фокус''.

\section{Критические точки ранга 2}  Тип критической точки $x_0$ ранга 2 в интегрируемой системе с тремя степенями свободы вычисляется следующим образом. Необходимо указать первый интеграл $F$, такой, что ${dF(x_0)=0}$ и ${dF\ne 0}$ в окрестности этой точки. Тогда, в частности, точка $x_0$ оказывается неподвижной для гамильтонова поля $\sgrad F$ и можно вычислить линеаризацию этого поля в точке $x_0$ --- симплектический оператор $A_F$ в шестимерном касательном пространстве к фазовому пространству в точке $x_0$. Этот оператор будет иметь четыре нулевых собственных числа, оставшийся сомножитель характеристического многочлена имеет вид ${\mu^2-C_F}$. При ${C_F<0}$ получим точку типа ``центр'' (соответствующий двумерный тор~-- эллиптический, является устойчивым многообразием в фазовом пространстве, пределом концентрического семейства трехмерных регулярных торов), а при ${C_F>0}$ получим точку типа ``седло'' (соответствующий двумерный тор~-- гиперболический, существуют движения, асимптотические к этому тору, лежащие на трехмерных сепаратрисных поверхностях).

В нашей задаче ситуация осложнена тем, что фазовое пространство задано в $\mathbb{R}^9$ тремя неявными уравнениями \eqref{nq1_1} и вычислять ограничения операторов на касательные пространства затруднительно. Однако функции в левых частях уравнений \eqref{nq1_1} служат функциями Казимира для естественного продолжения на $\mathbb{R}^9$ скобки Пуассона симплектической структуры пространства $T SO(3)$, поэтому при вычислении симплектических операторов вида $A_F$ они лишь добавят три нулевых корня в характеристический многочлен, имеющий в целом  девятую степень. Таким образом, мы заранее знаем, что при условии $\sgrad F=0$ искомый коэффициент $C_F$ есть коэффициент при $\mu^7$ в характеристическом многочлене $Z_F(\mu)$ оператора $A_F$ в $\mathbb{R}^9$:
$$
Z_F(\mu)=-\mu^7(\mu^2-C_F).
$$
Сам же оператор $A_F$ вычисляется и при наличии вырожденных скобок Пуассона (для рассматриваемого здесь пространства $\mathbb{R}^9$ они определены явно в \cite{BogRus2}). Отметим, что при вычислении характеристического многочлена через определитель соответствующей $(9{\times}9)$-матрицы трудности для некоторых функций оказываются слишком высоки даже при использовании мощных современных систем аналитических вычислений. Однако, заранее зная в данном случае структуру искомого многочлена, можем найти
\begin{equation*}
    C_F=\frac{1}{2}\left[\tr(A_F^2)-(\tr A_F)^2 \right]= \tr(A_F^2)/2,
\end{equation*}
так как след любого симплектического оператора равен нулю.

\begin{theorem}
{\it Тип критических точек ранга $2$ в критической подсистеме $\mathcal{M}_1$ определятся знаком квадрата характеристического показателя
\begin{equation}\label{nq1_12}
    \mu^2=\displaystyle{\frac{1}{2s}}\left[ 12 s^4-8(h-\frac{\lambda^2}{2})s^3+p^4-r^4\right]\left[ 2s^2-2(h+\frac{\lambda^2}{2})s+p^2\right],
\end{equation}
где $s$ -- значение параметра в соответствующей точке \eqref{nq1_3} поверхности $\Pi_1$. Критические точки имеют тип ``центр'' при ${\mu^2<0}$ и тип ``седло'' при ${\mu^2>0}$. При ${\mu=0}$ критические точки вырождены. На поверхности $\Pi_1$ множество вырожденных критических точек отвечает ребру возврата и линии касания с поверхностью $\Pi_2$.}
\end{theorem}
\begin{proof}
Рассмотрим функцию $L=2G-(p^2-\tau)H+s K$. Вычисляя ее косой градиент в точках \eqref{nq1_8} и приравнивая его к нулю, найдем
\begin{equation}\label{nq1_13}
    s=S_1, \qquad \tau = T_1 = p^2+2 S_1( S_1 - H+\frac{\lambda^2}{2}).
\end{equation}
Подчеркнем, что подстановки этих значений $s,\tau$ нужно выполнить {\it после} вычисления $\sgrad L$. Поэтому полученное поле $\sgrad L$ обращается в нуль лишь в самой рассматриваемой точке, но не в ее окрестности, следовательно, основное требование к интегралу $L$ выполнено. Здесь и далее при проверке тождеств либо соотношений, подобных равенствам \eqref{nq1_13}, используются представления \eqref{nq1_7}.

Вычислим $\tr(A_L^2)/2$, подставим значения \eqref{nq1_13}, затем -- \eqref{nq1_7} и \eqref{nq1_9}, и, в последнюю очередь, учтем уравнения \eqref{nq1_8}. Найденное выражение через исходные фазовые переменные средствами компьютерной алгебры разлагается на множители, которые дают искомое значение \eqref{nq1_12}. Техника преобразований здесь фактически совпадает с той, которая применялась в \cite{KhND07} при вычислении скобок Пуассона инвариантных соотношений, определяющих критические подсистемы. Характер критических точек определяется знаком величины \eqref{nq1_12}. Тот факт, что при равенстве ее нулю критические точки вырождены (т.е. нельзя указать другого интеграла с ненулевым характеристическим значением), вытекает из того, что коэффициент при $G$ в функции $L$ отличен от нуля, а равенство ${dG=0}$ приводит к критическим точкам ранга 1. Связь множества ${\mu^2=0}$ с геометрическими свойствами поверхностей $\Pi_{1,2}$ проверяется непосредственно. \end{proof}

\begin{theorem}
{\it Тип критических точек ранга $2$ в критической подсистеме $\mathcal{M}_2$ определятся знаком квадрата характеристического показателя
\begin{equation}\label{nq1_14}
    \mu^2=-\displaystyle{\frac{1}{s}}\left[ 8 \lambda^2 s^3 -r^4\right]\left[ 2s^2-2(h+\frac{\lambda^2}{2})s+p^2\right],
\end{equation}
где $s$ -- значение параметра в соответствующей точке \eqref{nq1_4} поверхности $\Pi_2$. Критические точки имеют тип ``центр'' при ${\mu^2<0}$ и тип ``седло'' при ${\mu^2>0}$. При ${\mu=0}$ критические точки вырождены. На поверхности $\Pi_2$ множество вырожденных критических точек отвечает ребру возврата и линии касания с поверхностью $\Pi_1$.}
\end{theorem}

Для доказательства снова возьмем функцию $L$, вычислим ее косой градиент и убедимся, что он обращается в нуль на многообразии $\mathcal{M}_2$, т.е. в силу уравнений \eqref{nq1_9}, но уже в подстановке
\begin{equation*}
    s=S_2, \qquad \tau = T_2 = 2\lambda^2 S_2.
\end{equation*}
Вычисляя матрицу линеаризации поля $\sgrad L$ и находя след ее квадрата, исключим переменные $y_1,y_2$ с помощью уравнений \eqref{nq1_10}. Получим искомое выражение \eqref{nq1_14}. Дальнейшие рассуждения такие же, как и в доказательстве теоремы~1.

\section{Критические точки ранга 1}Для вычисления типа критической точки $x_0$ ранга 1 нужно найти два интеграла $F_1,F_2$, независимых в проколотой окрестности $x_0$, таких, что в самой точке ${dF_1=0}$, ${dF_2=0}$, симплектические операторы $A_{F_1}$, $A_{F_2}$ независимы и существует их линейная комбинация, имеющая четыре различных собственных значения. Здесь и далее под дифференциалом функции, заданной на всем пространстве $\mathbb{R}^9$, понимается ограничение полного дифференциала на касательное пространство к фазовому пространству $\mathcal{P}^6$, заданному геометрическими интегралами. Простые уравнения, позволяющие проверять условие ${dF=0}$ на $\mathcal{P}^6$ не вводя неопределенных  множителей для ограничений \eqref{nq1_1}, приведены, например, в~\cite{Kh37}.

Как показано в \cite{Kh37}, особые периодические движения, заданные уравнениями (ОПД), принадлежат пересечению критических подсистем $\mathcal{M}_1,\mathcal{M}_2$. Фиксируем точку $x_0$ на таком решении и пусть
\begin{equation}\notag
    s_i=S_i(x_0), \qquad \tau_i = T_i(x_0), \qquad L_i(x)=L(x)|_{s=s_i, \tau=\tau_i} \qquad (i=1,2).
\end{equation}
Как следует из приведенных выше результатов, дифференциалы функций $L_1,L_2$ в точке $x_0$ равны нулю и отличны от нуля в близких к ней точках. Симплектические операторы, порожденные функциями $L_1,L_2$ в точке $x_0$, в разложении по собственным подпространствам будут отличны от нуля лишь на двумерных подпространствах в $T_{x_0}\mathcal{P}^6$, трансверсальных, соответственно, к многообразиям $\mathcal{M}_1,\mathcal{M}_2$, и иметь на этих подпространствах собственные числа $\pm \mu_1$, $\pm \mu_2$, определенные равенствами \eqref{nq1_12} с ${s=s_1}$ и \eqref{nq1_14} с ${s=s_2}$. Если ни одно из чисел $\mu_1,\mu_2$ не равно нулю, то поверхности $\Pi_1, \Pi_2$ пересекаются трансверсально, упомянутые двумерные подпространства образуют прямую сумму, и потому найдется линейная комбинация функций $L_1,L_2$, симплектический оператор которой имеет все различные собственные числа. Хотя это рассуждение и не совсем строго, но его можно подтвердить прямым вычислением.
Таким образом, если $J(x_0)$ принадлежит трансверсальному пересечению бифуркационных поверхностей, то точка $x_0$ является невырожденной.

\begin{theorem}
{\it Критические точки ранга $1$, лежащие на траекториях особых периодических движений, являются невырожденными за исключением случаев, когда точка в $\mathbb{R}^3$, образованная значениями первых интегралов, лежит в одном из следующих подмножеств:

{\rm 1)} ребре возврата поверхности $\Pi_1$, заданном на ней уравнением
\begin{equation}\label{nq1_15}
    12 s^4-8(h-\frac{\lambda^2}{2})s^3+p^4-r^4=0;
\end{equation}

{\rm 2)} ребре возврата поверхности $\Pi_2$, заданном на ней уравнением
\begin{equation}\label{nq1_16}
    8 \lambda^2 s^3 -r^4=0;
\end{equation}

{\rm 3)} кривой касания поверхностей $\Pi_1,\Pi_2$, заданной на них одним и тем же уравнением}
\begin{equation}\label{nq1_17}
    2s^2-2(h+\frac{\lambda^2}{2})s+p^2=0.
\end{equation}
\end{theorem}

Теперь для определения типа невырожденной критической точки ранга~1, т.е. такой точки $x_0$, что $J(x_0)$ не удовлетворяет ни одному из уравнений \eqref{nq1_15}--\eqref{nq1_17}, достаточно указать хотя бы один интеграл, у которого косой градиент в точке $x_0$ равен нулю, а соответствующий симплектический оператор имеет четыре ненулевых собственных значения.
В качестве такого интеграла можно взять функцию $K-2\sigma H,$ послужившую в \cite{Kh37} исходной для построения множества особых периодических движений.

\begin{theorem}
{\it Пусть фиксирован параметр $\lambda$, а константы $\sigma,u$ удовлетворяют уравнению \eqref{nq1_11}. Тогда тип невырожденной критической точки, заданной уравнениями {\rm (ОПД)}, определяется знаками следующей пары квадратов характеристических показателей {\rm (}собственных чисел симплектического оператора, порожденного интегралом $K-2\sigma H${\rm ):}}
\begin{equation}\label{nq1_18}
    \begin{array}{l}
      \displaystyle{ \nu_1^2= -\frac{4(u^3+r^4\lambda^4\sigma^3)}{u \lambda^2 \sigma(\lambda^2+\sigma)}; } \\
      \displaystyle{ \nu_2^2= \frac{u^2(\lambda^2+\sigma)^2-r^4\lambda^4\sigma^2}{u^4 \lambda^2 \sigma(\lambda^2+\sigma)}\left[ u^3\lambda^2-u^2\sigma(\lambda^2+\sigma)^2(\lambda^2+4\sigma)+r^4\lambda^6\sigma^3\right].}
    \end{array}
\end{equation}
\end{theorem}
Доказательство проводится непосредственным вычислением.

Отметим, что в точках особых периодических движений значения частных интегралов $S_1,S_2$ имеют достаточно простой вид \cite{Kh37}:
\begin{equation*}
    s_1=\frac{r^4\lambda^4\sigma^2-u^2(\lambda^2+\sigma)^2}{2u^2\lambda^2}, \qquad s_2=-\frac{u}{2\lambda^2\sigma}.
\end{equation*}
Применение компьютерной системы аналитических вычислений показывает, что обращение в нуль какой-либо из величин \eqref{nq1_18} влечет, в силу уравнения \eqref{nq1_11}, выполнение одного из условий \eqref{nq1_15}--\eqref{nq1_17}. Это еще раз подчеркивает связь вырожденных точек и геометрических особенностей бифуркационных поверхностей.
По сказанному выше, вычислять показатели типа \eqref{nq1_18} еще для одного интеграла нет необходимости.

Описание типов точек на траекториях маятниковых движений \eqref{nq1_5}, анонсированное в \cite{RyabConf}, получим, заметив, что
траектории семейств $\mathcal{M}_{\pm}$ состоят из критических точек первых интегралов
\begin{equation*}
G_{\pm}=\displaystyle{ \frac{2}{(a \mp b)\sqrt{ab}}(G \mp a b H)}.
\end{equation*}
Здесь постоянные множители введены для упрощения выражений собственных чисел.
\begin{theorem}
{\it Критические точки семейства $\mathcal{M}_-$ невырождены для всех значений энергии, кроме
\begin{equation*}
\displaystyle{h}=\displaystyle{\frac{(a+b)^4+(a- b)^2 \lambda^4}{2(a+b)^2 \lambda^2}}.
\end{equation*}
Тип этих точек определяется знаками пары квадратов собственных чисел симплектического оператора интеграла $G_-$:
\begin{equation*}
\displaystyle{\nu_{1,2}^2=- h +\displaystyle{\frac{\lambda^2}{2}}-\frac{4 ab \lambda^2}{(a+b)^2}  \pm \displaystyle{ \sqrt{\left(h-\frac{\lambda^2}{2}\right)^2+4 a b}}}.
\end{equation*}

Критические точки семейства $\mathcal{M}_+$ невырождены для всех значений энергии, кроме
\begin{equation*}
\displaystyle{h}=\displaystyle{\frac{\lambda^2}{2}} \pm 2\sqrt{a b}, \qquad \displaystyle{h}=\displaystyle{\frac{(a-b)^4+(a+b)^2 \lambda^4}{2(a-b)^2 \lambda^2}}.
\end{equation*}
Тип этих точек определяется знаками пары квадратов собственных чисел симплектического оператора интеграла $G_+$:
\begin{equation*}
\displaystyle{\nu_{1,2}^2= h-\displaystyle{\frac{\lambda^2}{2}}-\frac{4 a b\lambda^2}{(a-b)^2} \pm \displaystyle{ \sqrt{\left(h-\frac{\lambda^2}{2}\right)^2-4 a b}}}.
\end{equation*}
В частности, на интервале значений энергии \eqref{nq1_6} оператор имеет четыре комплексных собственных числа, поэтому соответствующие критические точки ранга $1$ на траекториях $\mathcal{M}_+$ имеют тип ``фокус-фокус''.}
\end{theorem}

\section{Критические точки ранга 0}Особенность нулевого ранга предполагает, в частности, равенство
$dH=0,$ что возможно лишь в неподвижных точках гамильтоновой
системы. Здесь их ровно четыре:
\begin{equation}\label{nq1_19}
c_k: \; {\boldsymbol\omega}={\boldsymbol 0}, \quad  {\boldsymbol
\alpha} = (\varepsilon_1  a ,\, 0,\, 0), \quad {\boldsymbol \beta}=
(0,\, \varepsilon_2 b, \, 0),
\end{equation}
где $\varepsilon_1^2=\varepsilon_2^2=1$ и $k=1,\ldots,4.$
Упорядочим $c_k$ по возрастанию $h$:
\begin{equation*}
    h_1=-a-b, \quad h_2=-a+b, \quad h_3=a-b, \quad h_4=a+b.
\end{equation*}
Тогда $c_1,c_4 \in \mathcal{M}_+$, $c_2,c_3 \in \mathcal{M}_-$. Точки \eqref{nq1_19} принадлежат обеим критическим подсистемам $\mathcal{M}_1,\mathcal{M}_2$ за одним исключением, а именно, точка $c_4$ самого верхнего положения равновесия не принадлежит $\mathcal{M}_1$, если $h=a+b$ лежит в интервале \eqref{nq1_6}.

Как показано в работе \cite{ZotKh}, индекс Морса гамильтониана $H$ в точке $c_k$ равен ${k-1}$, что в значительной мере определяет поведение системы в ее окрестности. В частности, отсюда сразу следует, что лишь самое нижнее положение равновесия устойчиво. Однако исследование характера неустойчивости остальных точек и строгая классификация требуют вычисления типа этих точек как критических точек интегрального отображения.

\begin{theorem}
{\it Точки $c_{1,2}$ для всех значений $\lambda \geqslant 0$ являются невырожденными критическими точками ранга~$0$ интегрального отображения. При этом $c_1$ имеет тип {\rm ``}центр-центр-центр{\rm '',} $c_2$\,--- тип {\rm ``}центр-центр-седло{\rm ''.} Точка $c_3$ при всех $\lambda \geqslant 0$ имеет тип {\rm ``}центр-седло-седло{\rm '',} однако, размерность алгебры симплектических операторов в этой точке падает до двух при $\lambda^2=q_-$, где
$$
q_-=\frac{(a+b)^2}{a-b}.
$$
Точка $c_4$ является невырожденной при значениях $\lambda^2 \notin \{q_+,q_1,q_2\}$, где
\begin{equation}\label{nq1_20}
\begin{array}{l}
q_+=\displaystyle{\frac{(a-b)^2}{a+b}}, \quad q_1=2 \left(\sqrt{\mathstrut a} -\sqrt{\mathstrut b}\right)^2 < q_2= 2 \left(\sqrt{\mathstrut a} +\sqrt{\mathstrut b}\right)^2.
\end{array}
\end{equation}
Она имеет тип {\rm ``}седло-седло-седло{\rm ''} при $\lambda^2 < q_1 $, {\rm ``}седло-фокус-фокус{\rm''} при $\lambda^2 \in (q_1,q_2)$, и {\rm ``}седло-центр-центр{\rm ''} при $\lambda^2 > q_2$, значение $\lambda^2=q_+$ отвечает падению размерности алгебры операторов и на тип не влияет.}
\end{theorem}
\begin{proof}
Поскольку все точки $c_k$ являются невырожденными критическими точками гамильтониана в смысле Морса, то в этих точках также ${dK=0}, {dG=0}$. Независимость операторов $A_H, A_K, A_G$ в точках \eqref{nq1_19} проверяется непосредственно в пространстве $\mathbb{R}^9$ и нарушается лишь для точек $c_{3,4}$ при значениях $\lambda^2 = q_\mp$ соответственно. Одновременно иметь по паре совпадающих собственных чисел эти три оператора могут лишь в точке $c_4$ при ${\lambda^2=q_{1,2}}$, где $q_1,q_2$ определены в \eqref{nq1_20}.
Переменную характеристического многочлена оператора $A_H$ в точке $c_k \in \mathbb{R}^9$ обозначим через $\varkappa$. После сокращения на множитель $\varkappa^3$, отвечающий геометрическим интегралам, этот многочлен имеет следующие корни относительно~$\varkappa^2$:
\begin{equation}\notag
\begin{array}{l}
\varkappa_1^2=h_k,\qquad \varkappa_{2,3}^2 = \displaystyle{ \frac{1}{4} \left[ h_k-\frac{\lambda^2}{2} \pm \sqrt{\left(h_k-\frac{\lambda^2}{2}\right)^2-4 \varepsilon_1 \varepsilon_2 a b} \, \right]}.
\end{array}
\end{equation}
Поэтому $A_H$ не является регулярным элементом соответствующей подалгебры лишь для точки $c_4$ при одном из условий ${\lambda^2=q_{1,2}}$. Утверждение теоремы следует теперь из анализа знаков найденных корней. \end{proof}

\section*{Заключение}\ В работе представлена вся информация об аналитических характеристиках критических точек интегрального отображения для неприводимой интегрируемой системы с тремя степенями свободы, описывающей движение гиростата типа Ковалевской в двойном силовом поле. С использованием условий существования критических движений, найденных в \cite{KhIISmir}, предоставляется возможность построения диаграмм критических подсистем, т.е. таких множеств на плоскости констант двух независимых (почти всюду) первых интегралов, которые классифицируют критические точки по рангам и типам. После этого в большинстве случаев можно указать и ха\-рактер бифуркации интегрального многообразия, происходящей при пересечении соответствующего участка бифуркационной поверхности в пространстве $\mathbb{R}^3$ констант общих интегралов. В силу высокой технической сложности задачи вряд ли можно рассчитывать на получение полной аналитической классификации топологических инвариантов (т.е. построения всех разделяющих поверхностей с указанием условий существования движений в пространстве всех параметров задачи), однако с помощью компьютерного моделирования можно построить новые интересные примеры трехмерной фазовой топологии.

\begin{abstract}[en]
\title{Types of critical points of the Kowalevski gyrostat in double field}
\author{M.P.~Kharlamov, P.E.~Ryabov, A.Y.~Savushkin, G.E.~Smirnov}
The problem of motion of the Kowalevski type gyrostat in double force field is considered.
According to the classification used in the theory of Liouville integrable Hamiltonian systems, the types of critical points of all ranks of the integral map are calculated.
\end{abstract}

\begin{abstract}
\title{Типи критичних точок гіростата Ковалевської в подвійному полі}
\author{М.П.~Харламов, П.Є.~Рябов, О.Ю.~Савушкін, Г.Є.~Смірнов}
Відповідно до  класифікації, прийнятої в теорії інтегровних за Ліувіллем гамільтонових систем, обчислено типи критичних точок всіх рангів інтегрального відображення задачі про рух гіростата Ковалевської у подвійному силовому полі.
\end{abstract}

\makeaddressline

\end{document}